\documentclass[11pt, oneside]{article}   	% use "amsart" instead of "article" for AMSLaTeX format
\usepackage{geometry}                		% See geometry.pdf to learn the layout options. There are lots.
\usepackage{graphicx}				% Use pdf, png, jpg, or eps§ with pdflatex; use eps in DVI mode
\usepackage{amssymb}
\usepackage{amsmath}
\usepackage{color}
\usepackage{animate}
\usepackage{soul}
\usepackage{subcaption}
\usepackage{authblk}

\usepackage{graphicx}
\usepackage{cancel}
\usepackage{multicol}

\title{Aneurysm Growth Time Series Reconstruction Using Physics-informed Autoencoder}
\author{Jiacheng Wu}
\affil{ University of California Berkeley}
\date{}							% Activate to display a given date or no date
\begin{document}
\maketitle

\section*{Abstract}
\begin{figure}[h]
        \centering
        \includegraphics[width=0.6\textwidth]{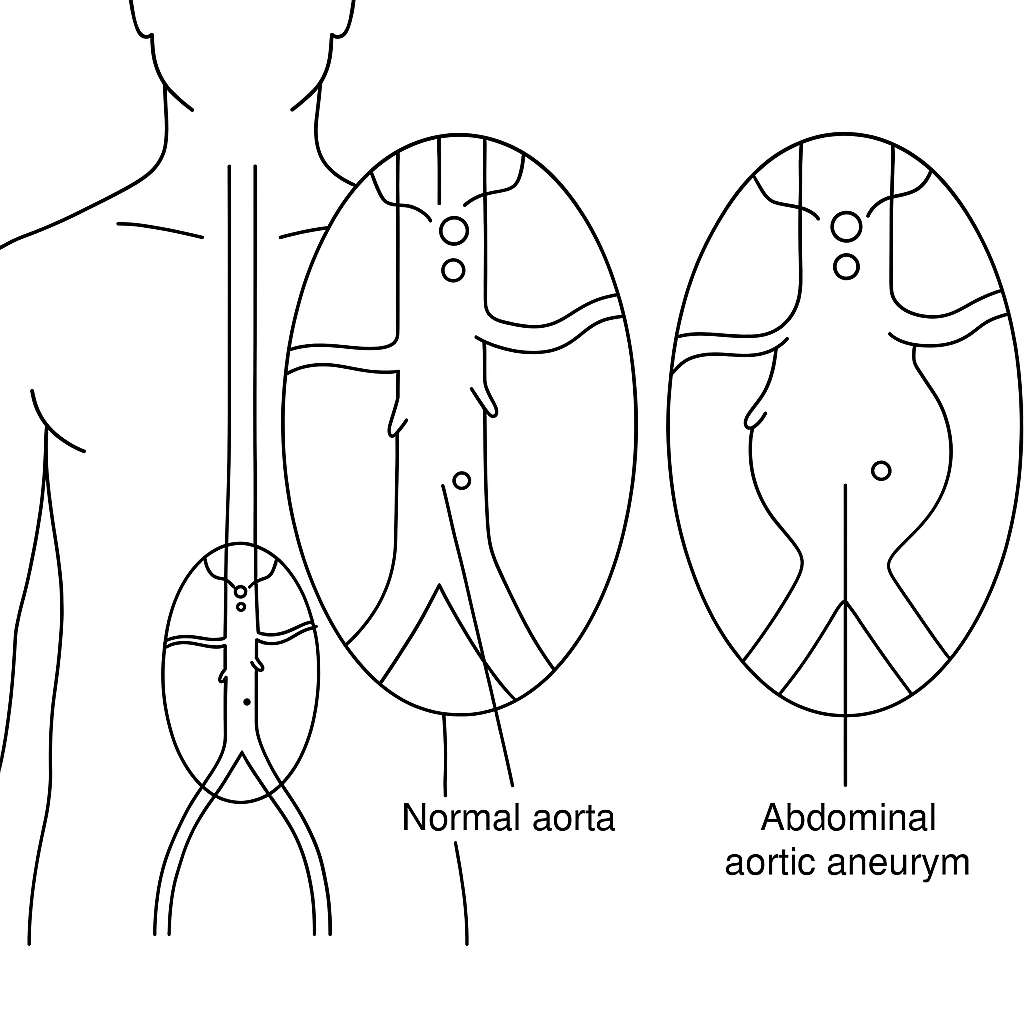}
        \caption{Aneurysm is a local expansion of arterial wall~\cite{cronenwett1985actuarial}. }
        \label{fig:aneurysm}
    \end{figure}

Arterial aneurysm (Fig.~\ref{fig:aneurysm}) is a bulb-shape local expansion of human arteries, the rupture of which is a leading cause of morbidity and mortality in US. Therefore, the prediction of arterial aneurysm rupture is of great significance for aneurysm management and treatment selection. The prediction of aneurysm rupture depends on the analysis of the time series of aneurysm growth history. However, due to the long time scale of aneurysm growth, the time series of aneurysm growth is not always accessible. We here proposed a method to reconstruct the aneurysm growth time series directly from patient parameters. The prediction is based on data pairs of [patient parameters, patient aneurysm growth time history]. To obtain the mapping from patient parameters to patient aneurysm growth time history, we first apply  autoencoder to obtain a compact representation of the time series for each patient. Then a mapping is learned from patient parameters to the corresponding compact representation of time series via a five-layer neural network. Moving average and convolutional output layer are implemented to explicitly taking account the time dependency of the time series. 

Apart from that, we also propose to use prior knowledge about the mechanism of aneurysm growth to improve the time series reconstruction results. The prior physics-based knowledge is  incorporated as constraints for the optimization problem associated with autoencoder. The model can handle both algebraic and differential constraints. Our results show that including physical model information about the data will not significantly improve the time series reconstruction results if the training data is error-free. However, in the case of training data with noise and bias error, incorporating physical model constraints can significantly improve the predicted time series. 

%[TODO: Add figures to explain what is normal vessel and vessel with aneurysm]
%[TODO: Describe the design for each network and how many layers and how many neurons in each layer]
%[TODO: Why need to combine machine learning with physical model informations]

%\subsection{}
\section{Introduction}
Modeling the progression of arterial aneurysm growth has always been a key part for cardiovascular research as it is a leading cause of morbidity and mortality in US. Physical models derived from physical laws and empirical relations derived from experiments or prior knowledge have been used to capture the behaviors in aneurysm progression. For example, \cite{humphrey2002constrained} proposed a theory for vascular growth and remodeling
(G\&R) based on the constrained mixture theory, which provides the fundational framework to study aneurysm growth. \cite{wu2015coupled} proposed a computational framework that couples vascular (G\&R) with hemodynamics to simulate aneurysm progression. In general, physics-based models, once established, can give predictions in a relatively large range without much uncertainty and randomness. However, biological systems are by nature to be a very complex system, which makes it impossible to model every part of the problem using physical models. 

In recent years, due to the availability of bio-sensors and computation powers, date-driven machine learning approaches have been widely used to give predictions in biomedical applications. The significant progress in deep learning recently has been pushing this trend even further, and has shown promising leads in fields like computer vision and autonomous driving. However, due to the uniqueness of biomedical research and applications, a purely data-driven approaches still possess several weaknesses:
\begin{itemize}
\item {\bf Available data sets are limited in biomedical applications.} A key factor for the current data-driven machine learning techniques to perform well is the availability of large data sets. However, due to reasons like the high expenses and invasiveness of clinical measurements, and the relative long term evolution process of diseases, large data sets are not always available in biomedical applications.
\item {\bf Prior knowledge exists.} While data may be limited, there exists a large body of physical models that can provide prior informations about disease progression. Therefore, ideally, there is no need to learn everything from scratch.
\item {\bf Interpretability and trustability.} Compared to other fields, biomedical problems require not only accuracy but also interpretability. As this is directly related  to human life, physicians who make decisions on treatments not only needs an accurate prediction, but also want the physiological reasons behind these predictions. Back-box type data-driven approaches usually cannot provide meaningful insights to the physics behind the predictions. Whether the framework can provide a physiological reasons behind predictions will directly influence the trustability of the results.
\end{itemize}
All the above reasons motivates us to use an prediction approach that can combine machine learning with physical models.
Along this direction, \cite{willard2020integrating} presented a general overview of integrating traditional physics-based modeling with machine learning techniques, \cite{wu2019adding} proposed a noval method to add algebraic constraints into generic data filtering problemes using likelihood functions and \cite{raissi2019physics} proposed a framework to train neural networks that respects physics described by differential equations.
 Here we use the prediction of arterial aneurysm growth time series to implement our physics-informed time series learning for the purpose of proof-of-concept. However, the application of this framework will not be limited in this specific field. 

\section{Method}
\subsection{Problem definition}
The time evolution of arterial aneurysm growth is described by the time series of vascular states (defined as ${X}(t)$). This is very important for determining the long term response of vascular tissue to biomechanical (or biochemical) stimulus, and therefore determining the rupture of the aneurysms. An aneurysm with unstably fast growing speed is likely to rupture. The time series for each individual patient are determined by a set of patient-specific physical parameters $\theta$ that can easily obtain from clinical measurements, such as arterial stiffness, vascular homeostatic stress etc. The goal is to construct the mapping $\mathcal{F}: \theta \mapsto {X}$ using machine learning approach.

The time series of arterial aneurysm growth that capture patient-specific characteristics (geometry and stress) can be acquired by MR angiography \cite{boussel2008aneurysm}.  However, this data set is not immediately available right now due to expense of clinical measurements and long range of observation time (it takes years for aneurysms to grow). Therefore, for proof-of-concept we instead use synthetic data generated from a well-established model \cite{wu2016stability} rather than actual patients. The governing equations for the model is shown in (\ref{eq:state_equation}), with five physical states ${ X}(t)= \left(M(t), r(t), y(t), m(t), \sigma(t) \right)$, denoting mass density, vessel radius, generalized stiffness, mass production rate and vascular stress. By varying patient-specific parameters (the parameter space is uniformly sampled), we can generate a library of synthetic data relating different patient parameters $\theta$ with their observed time histories ${X}(t)$.  The data pairs $(\theta, { X})$ will be used for training.
\begin{equation} 
\theta = \left\{ k_g, \alpha, E, R\right\}
\Rightarrow
\left\{
  \begin{tabular}{ccc}
  $\dot{M}(t)=M(t)k_g\left[\sigma(t)-\sigma_h \right] $ \\
  $\dot{r}(t)=\frac{1}{k(t)}\left[\alpha r(t) -\frac{m(t)}{r(t)}k_1\right] $ \\
  $\dot{y}(t)=k_2 \frac{m(t)}{r^2(t)}-\alpha y(t) $ \\
  $m(t)=M(t)\left[k_g\left[\sigma(t)-\sigma_h\right] +f_h \right]$ \\
  $\sigma(t)=\frac{\rho P r(t)^2}{M(t) R}$ 
  \end{tabular}
\right \}
\Rightarrow
{X} =\left\{
  \begin{tabular}{ccc}
  $M_0, M_1 , ... , M_t, ... $ \\
  $r_0, r_1, ... , r_t, ...  $ \\
  $y_0, y_1, ... , y_t, ...  $ \\
  $m_0, m_1, ... , m_t, ...  $ \\
  $\sigma_0, \sigma_1, ... , \sigma_t, ...  $
  \end{tabular}
\right. 
\label{eq:state_equation}
\end{equation}
The other goal is to show physics-informed machine learning (machine learning + information about physical model) can perform better than solely data-driven counterpart. Here two scenarios  of learning are proposed: 
\begin{itemize}
\item Scenario 1: Only data pairs $(\theta, { X})$ are given.
\[
\mbox{Date pairs} \Rightarrow \mbox{the mapping} ~\mathcal{F}: \theta \mapsto { X}
\]
\item Scenario 2:  Apart from data pairs $(\theta, { X})$, part of the physical model is also known. 
\begin{equation}
\mbox{Date pairs} +
\left\{
  \begin{tabular}{ccc}
 \st{$\dot{M}(t)=M(t)k_g\left[\sigma(t)-\sigma_h \right] $} \\
  \st{$\dot{r}(t)=\frac{1}{k(t)}\left[\alpha r(t) -\frac{m(t)}{r(t)}k_1\right] $} \\
  \st{$\dot{y}(t)=k_2 \frac{m(t)}{r^2(t)}-\alpha y(t) $} \\
  $\textcolor{black}{m(t)=M(t)\left[k_g\left[\sigma(t)-\sigma_h\right] +f_h \right]}$ \\
  $\textcolor{black}{\sigma(t)=\frac{\rho P r(t)^2}{M(t) R}}$ 
  \end{tabular}
\right \}
\Rightarrow \mbox{the mapping}~ \mathcal{F}: \theta \mapsto { X} \nonumber
\end{equation}
\end{itemize}

\subsection{Compact representation of time series using autoencoder}
The mapping $\mathcal{F}$  is constructed by first using an autoencoder \cite{ng2011sparse} to obtain a compact representation $Z$ of the time series data ${ X}(t)$. This reformulates our model as a mapping from patient parameters $\theta$  to a compact representation $Z$  of the aneurysm growth process. This is motivated by the fact that physics governing the high-dimensional time series is usually embedded in a much lower dimension, and traditional representation of the dynamics  often contains unnecessary or redundant information.

\begin{figure}[!htb]
    \centering
%    \begin{minipage}{.5\textwidth}
%        \centering
%        \includegraphics[width=0.6\textwidth]{autoencoder_manifold}
%        \caption{Representation of data with low dimensional manifold.}
%        \label{fig:low_dim_manifold}
%    \end{minipage}%
    \begin{minipage}{1.0\textwidth}
        \centering
        \includegraphics[width=0.8\textwidth]{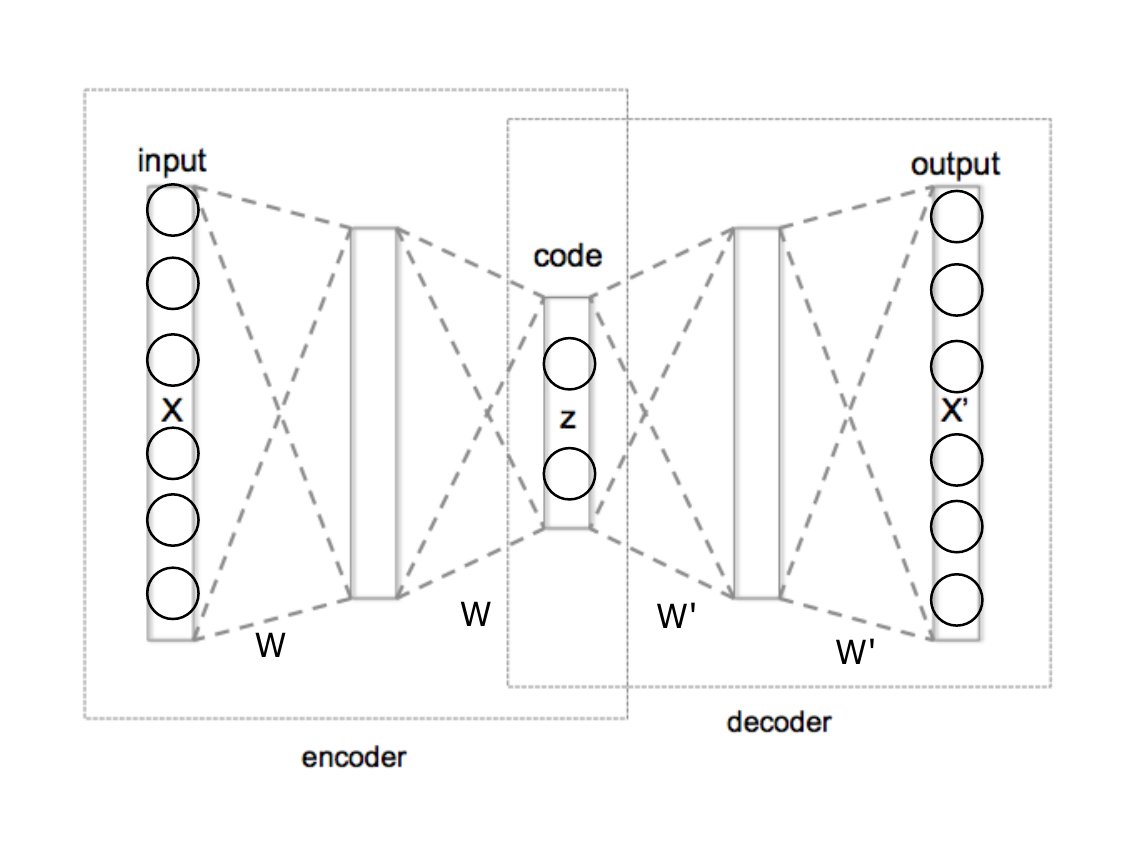}
        \caption{Neural networks for autoencoder}
        \label{fig:autoencoder}
    \end{minipage}
\end{figure}

The architecture  of the autoencoder is pictured in Fig.~\ref{fig:autoencoder} and can be mathematically defined by the encoding and decoding process
\begin{align}
Z &= \sigma(WX + b) ~~\mbox{encoder} \nonumber \\
X' & = \sigma'(W'Z + b') ~~\mbox{decoder} 
\end{align}
where $W, b, W', b'$ are the connection weights  and biases  of the autoencoder neural network, and $\sigma, \sigma'$ are the activation function for the neurons. $X$ is the original time series and $X'$ is the reconstructed time series. The latent state $Z$ is the compact representation for the corresponding time series. The encoder compresses the input data $X$ to the low dimensional representation $Z$ and the decoder reconstructs the time series $X'$ from $Z$. To make sure $Z$ is a good representation of $X$, we want the difference between rhe original data $X$ and the reconstructed one $X'$ to be as small as possible. This motivates the optimization problem for training an autoencoder:
\begin{align}
&\min_{W, b, W', b'} \sum_{i=1}^N \|X_i-X_i'\|^2  \label{eq:autoencoder_no_constraint} \nonumber \\
&s.t.~Z_i = \sigma(WX_i + b), ~X'_i  = \sigma'(W'Z_i + b') ,~\forall i \in \{1,2,...,N\}\;.
\end{align}
where $N$ is the total number of data points in the training set. 

%\begin{columns}
%\column{0.5\textwidth}
%\begin{figure}
%  \centering
%    \includegraphics[width=0.6\textwidth]{autoencoder_manifold}
%      \caption{\tiny Representation of data with low dimensional manifold.}
%\end{figure}
%\column{0.5\textwidth}
%\begin{figure}
%  \centering
%    \includegraphics[width=0.8\textwidth]{autoencoder_layer_1}
%      \caption{\tiny Neural networks for autoencoder}
%\end{figure}
%\end{columns}

\subsection{Incorporate physical model information as constraints for the learning optimization problems}
In the case of small data training problem, which is especially true for long term medical problems like aneurysm growth, prior knowledge about the physical models can be a strong augmentation to guide the learning process. These physical models are either derived from physical laws or from existed empirical knowledge. In general, they represent the relations between different physical variables in the form of algebraic equations or differential equations
\begin{equation}
f(X) = 0~\mbox{or}~\dot{X} = g(X)\;. \label{eq:constraints}
\end{equation}
These physical model informations are incorporated as constraints for the learning optimization problem (\ref{eq:autoencoder_no_constraint}). However, the optimization variables in (\ref{eq:autoencoder_no_constraint}) are autoencoder model parameters $W, b, W', b'$ while the physical constraints are on the physical variables. Therefore, we need to convert the constraints (\ref{eq:constraints}) onto  $W, b, W', b'$. This is done by enforcing constraints on the reconstructed data $X'$, as $X'$ is an explicit function of $W, b, W', b'$:
\begin{equation}
X' = \sigma'(W'(\sigma(WX+b))+b')\;.
\end{equation}
Therefore, by taking account the physical constraints, we constructed a new constrained optimization problem for the learning process
\begin{align}
&\min_{W, b, W', b'} \sum_{i=1}^N \|X_i- \sigma'(W'(\sigma(WX_i+b))+b')\|^2  \label{eq:autoencoder_no_constraint} \nonumber \\%~~\mbox{with}~Z_i = \sigma(WX_i + b), ~X'_i  = \sigma'(W'Z_i + b') \nonumber \\
s.t.~& f(X'_i(W, b, W', b')) = 0 \nonumber \\
& \frac{d}{dt}(X'_i(W, b, W', b')) = g(X'_i(W, b, W', b')), ~\forall i \in \{1,2,...,N\}\;.
\end{align}
For the differential equation constraints, we actually convert them into algebraic forms first before they are included. This is achieved by discretizing the time derivatives using Crank-Nicolson method~\cite{crank1947practical}:
\begin{equation}
\frac{X(t+\Delta t) - X(t)}{\Delta t} \approx \frac{1}{2}\left[ g(X(t+\Delta t)) + g(X(t))  \right]
\end{equation}
Define a shift forward operator for the time series as $\bf S$, then 
\begin{equation}
X(t+\Delta t)  = {\bf S} X(t)\;.
\end{equation}
In this way, the differential equation constraints are converted to:
\begin{equation}
\frac{ {\bf S} X(t) - X(t)}{\Delta t} \approx \frac{1}{2}\left[ g( {\bf S} X(t)) + g(X(t))  \right]
\end{equation}
which can be fit into the general algebraic constraint form
\begin{equation}
f( {\bf S} X, X) = 0\;.
\end{equation}
For proof of concept, the physical constraints we choose to implement in the five state equations (\ref{eq:state_equation}) are
\begin{align}
m(t)&=M(t)\left[k_g\left[\sigma(t)-\sigma_h\right] +f_h \right] \nonumber \\
\sigma(t) &=\frac{\rho P r(t)^2}{M(t) R}
\end{align}
where the first equation is an empirical assumption for vascular tissue growth and the second equation is a variant of Laplace Law, which is derived from Newton Second Law for ideal cylindrical geometry. While these physical constraints provide useful insights about the learning problem, they are only an approximation of the truth and may only be accurate enough under certain assumption. Therefore, we do not want these constraints to be strictly satisfied when implemented, and enforce them using penalty method~\cite{galar2013aggregation}
\begin{align}
\min_{W, b, W', b'} &\sum_{i=1}^N \|X_i- \sigma'(W'(\sigma(WX_i+b))+b')\|^2  \label{eq:autoencoder_no_constraint} \nonumber \\
&+ \sum_{c=1}^{N_c}\lambda_c\sum_{i=1}^N\| f_c(X'_i(W, b, W', b'))\|^2\;,
\end{align}
where $c$ denotes the indices for different constraints, and $\lambda_c >0$ are the corresponding penalty constants. By varying the magnitude of $\lambda_c$, we can control how strict the corresponding constraint needs to be enforced.

\subsection{Construct the mapping from patient-specific parameters to latent representation of time series}
\begin{figure}[h]
        \centering
        \includegraphics[width=0.95\textwidth]{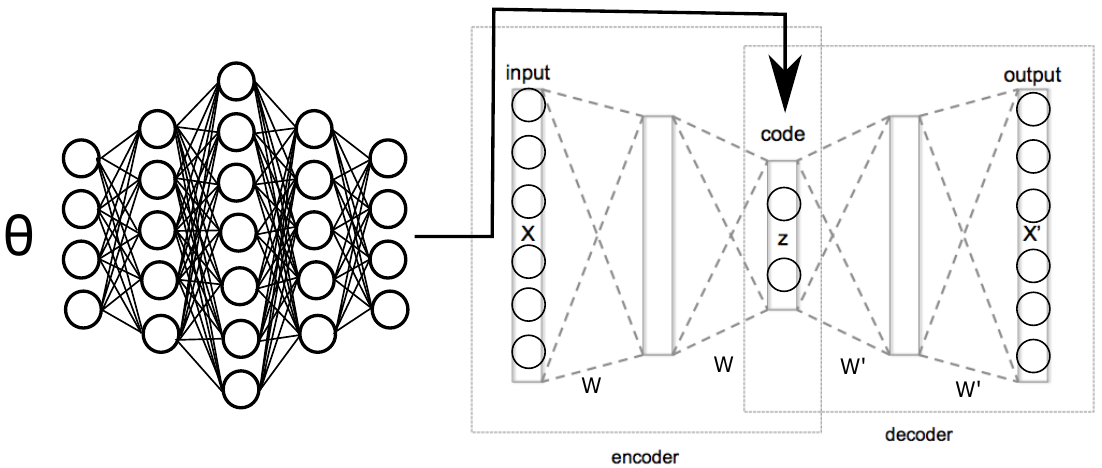}
        \caption{Neural networks mapping from patent-specific parameters $\theta$ to the latent representation $Z$ for the time series}
        \label{fig:theta_to_z}
\end{figure}

The autoencoder helps to extract a compact representation $Z$ for the time series of each individual patient. Ultimately, we want to reconstruct the time series from the corresponding patient parameters $\theta$. Therefore, if we can construct the mapping from $\theta$ to $Z$, then the time series can be reconstructed via the decoder (i.e. map from $Z$ to $X'$). The mechanism is shown in Fig.~\ref{fig:theta_to_z}. 
In this work, a neural network in the left part of Fig.~\ref{fig:theta_to_z} is used to construct the mapping:
\begin{equation}
\mathcal{G}: \theta \mapsto Z\;.
\end{equation}
The network consists one input layer, three hidden layers and one output layer. The number of neurons in each layer from input to output is equal to $4,8, 16, 10, 5$. The reason for choosing this network topology is to first obtain more features by combining different input features and then use the extended features to predict the final output. 

\subsection{Capture the time dependency of the time series: moving average and convolutional layer}
In the above learning framework, the time dependency of the time series are not directly modeled. It is taken account implicitly by autoencoder. However, sometimes we may want to explicitly model the time dependency within the time series. Here two potential ways are provided to explicitly model the time dependency. First we start from the differential equations (\ref{eq:state_equation}) generating the time series data:
\begin{equation}
\dot{X} = g(X)
\end{equation}
If we use Forward Euler method to discretize the time derivative, then the current state $X(t)$ can be expressed as 
\begin{equation}
X(t) = X(t-1) + F(X(t-1))dt
\end{equation} 
which means the current state should be a function of state in previous states. In a more general form
\begin{equation}
X_t = \mathcal{N}(X_t, X_{t-1}, X_{t-2}, ...)
\end{equation}
where the function $\mathcal{N}(\cdot)$ defines the way current state depend on the previous steps. In the case of that $\mathcal{N}(\cdot)$ is a linear function, the relation yields moving average method~\cite{devcic2006weighted}. For example, 
\begin{equation}
X_t = aX_{t-2} + bX_{t-1}+ cX_t
\end{equation}
where $a =\frac{1}{4}, b=\frac{1}{4} , c = \frac{1}{2}$. Ideally, implementing moving average method to the reconstructed time series should yield better reconstructed results (especially for systems with slow dynamics).

Another way to take account time dependency explicitly is through convolutional neural networks~\cite{lecun2015lenet}
\begin{equation}
X_t = \mathcal{A}(K*X)_t = \mathcal{A}( \sum_{h=0}^p K_h X_{t-h})
\end{equation}
where $\mathcal{A}(\cdot)$ is the activation function and $K$ is the kernel for the convolutional layer. The kernel $K$ will take account the local time dependency within the time series. 

\begin{figure}[!htb]
    \centering
    \begin{minipage}{1.0\textwidth}
        \centering
        \includegraphics[width=0.8\textwidth]{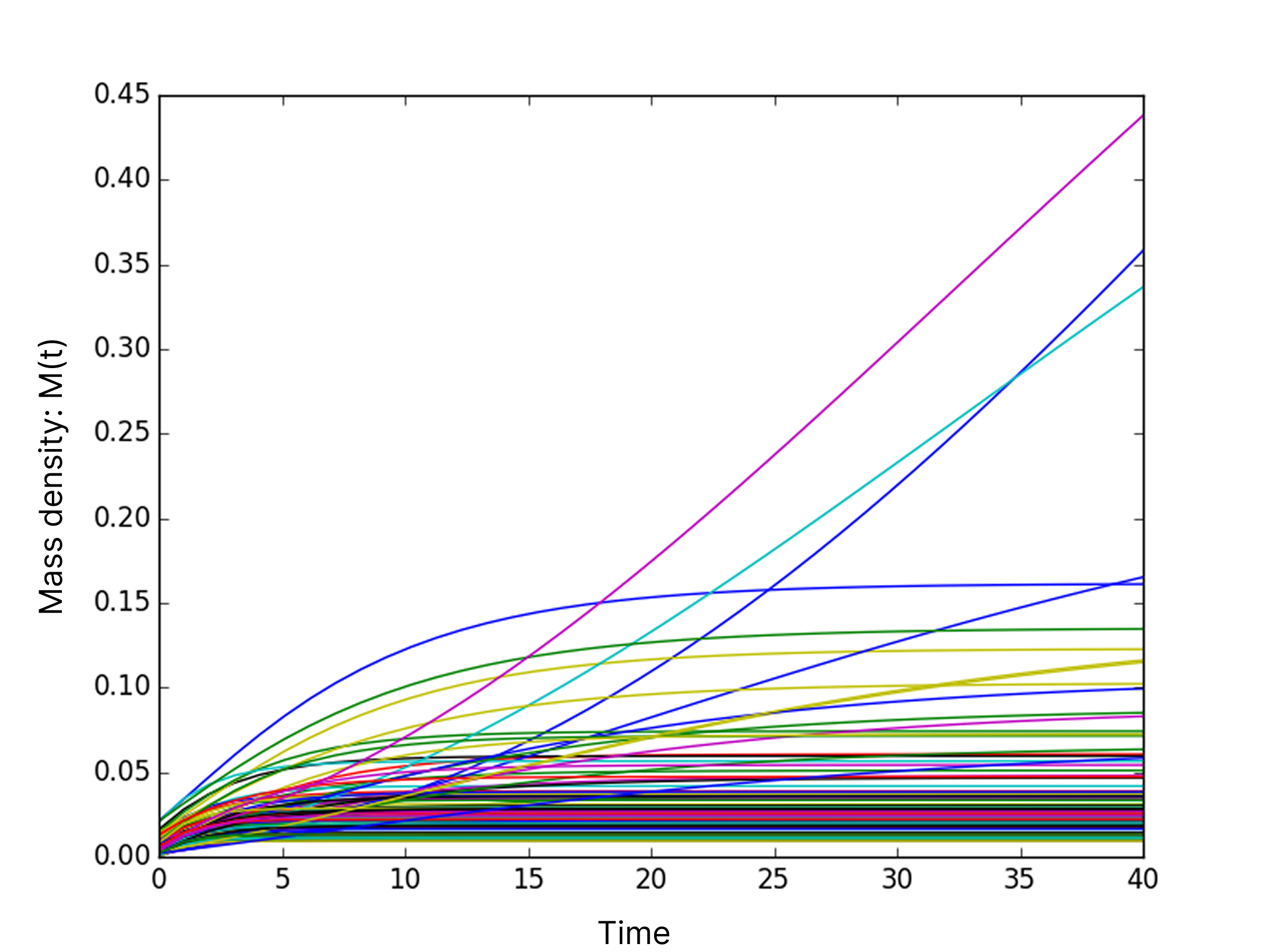}
        \caption{Original time series}
        \label{fig:orignal}
    \end{minipage}%
\end{figure}

\begin{figure}[!htb]
    \centering
    \begin{minipage}{1.0\textwidth}
        \centering
        \includegraphics[width=0.8\textwidth]{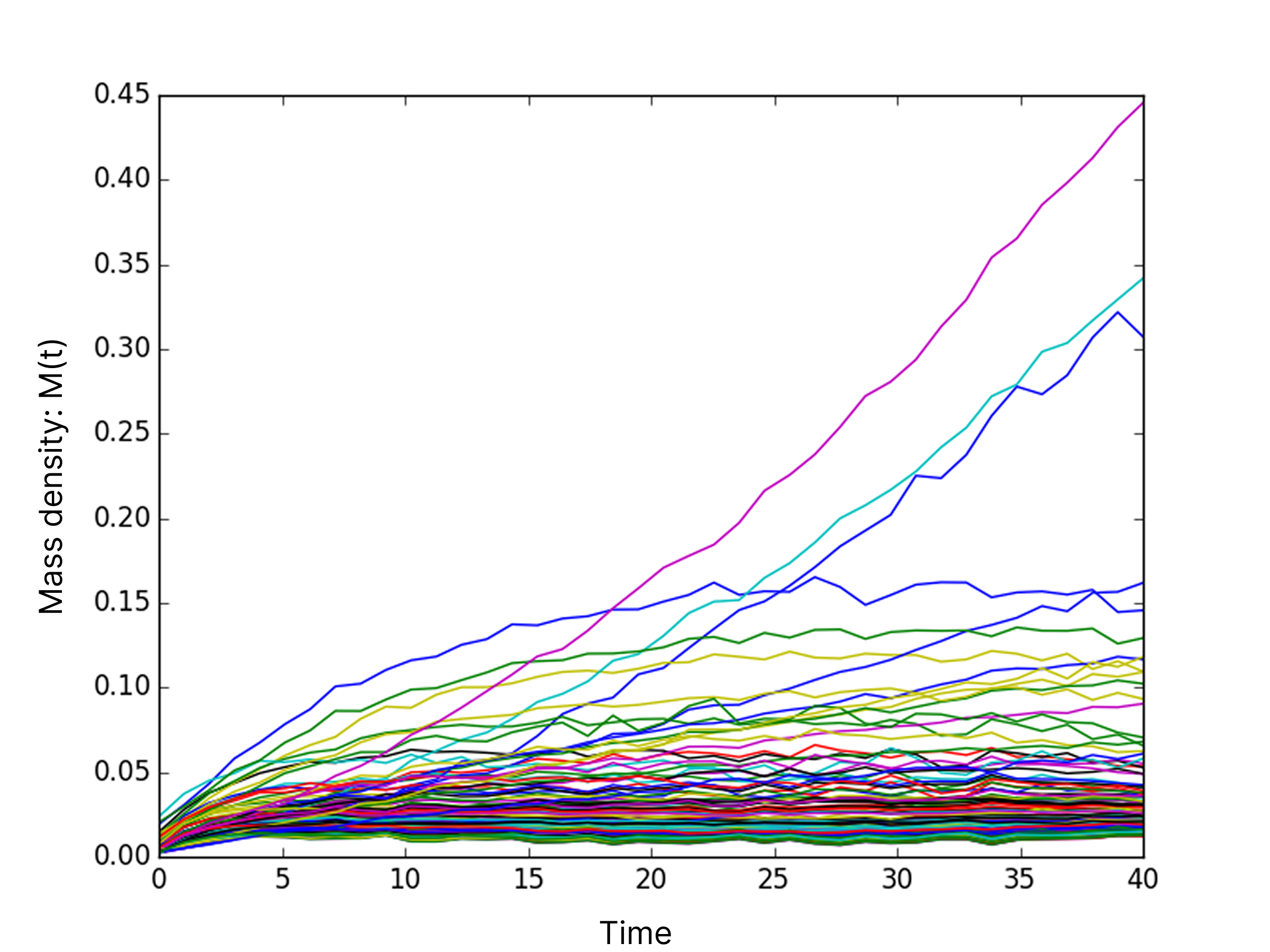}
        \caption{Reconstructed time series}
        \label{fig:reconstruct}
    \end{minipage}
\end{figure}

\begin{figure}[!htb]
    \centering
    \begin{minipage}{1.0\textwidth}
        \centering
        \includegraphics[width=0.8\textwidth]{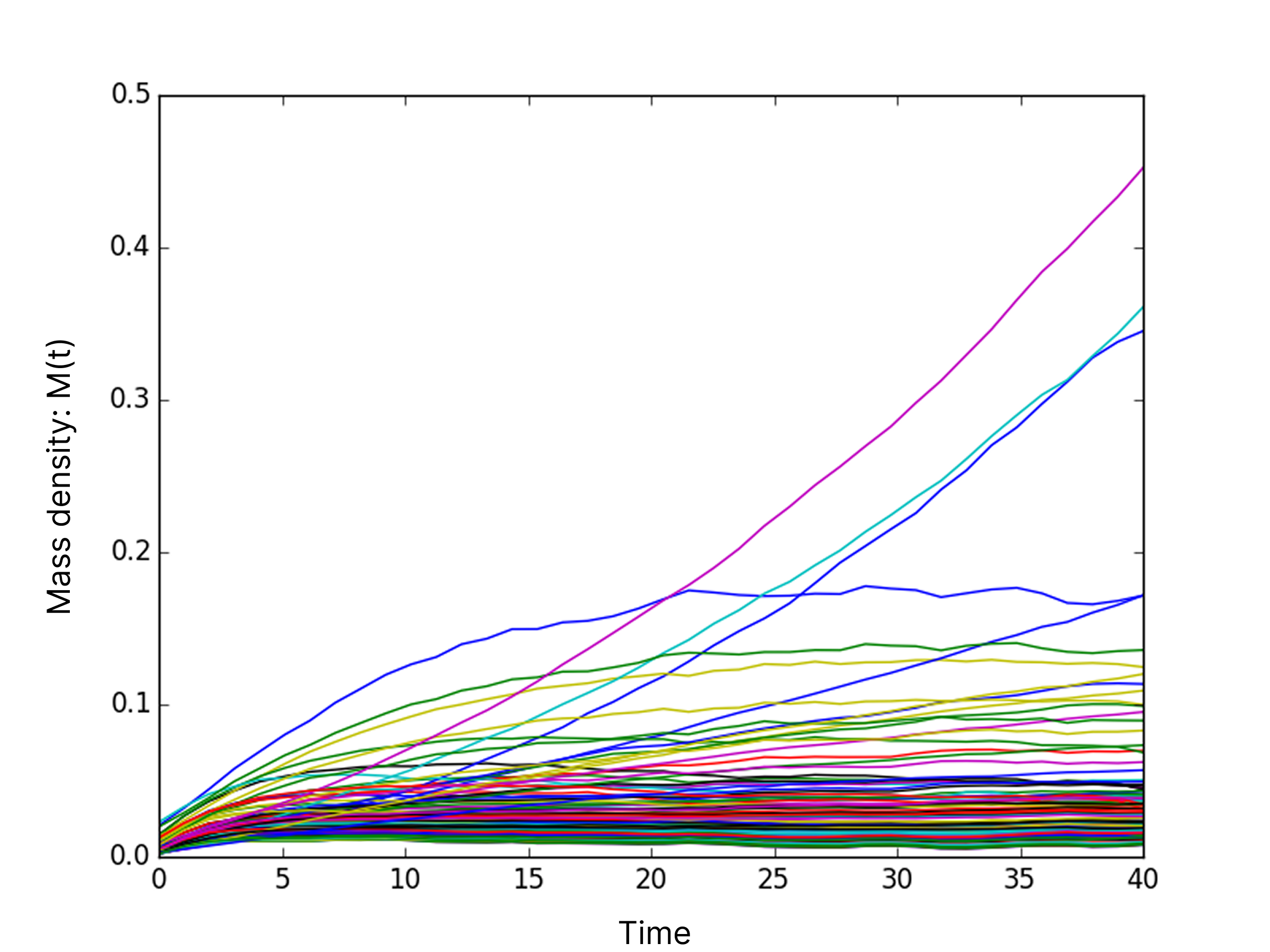}
        \caption{Reconstructed time series with moving average implemented}
        \label{fig:reconstruct_moving_average}
    \end{minipage}
\end{figure}

\begin{figure}[!htb]
    \centering
    \begin{minipage}{1.0\textwidth}
        \centering
        \includegraphics[width=0.8\textwidth]{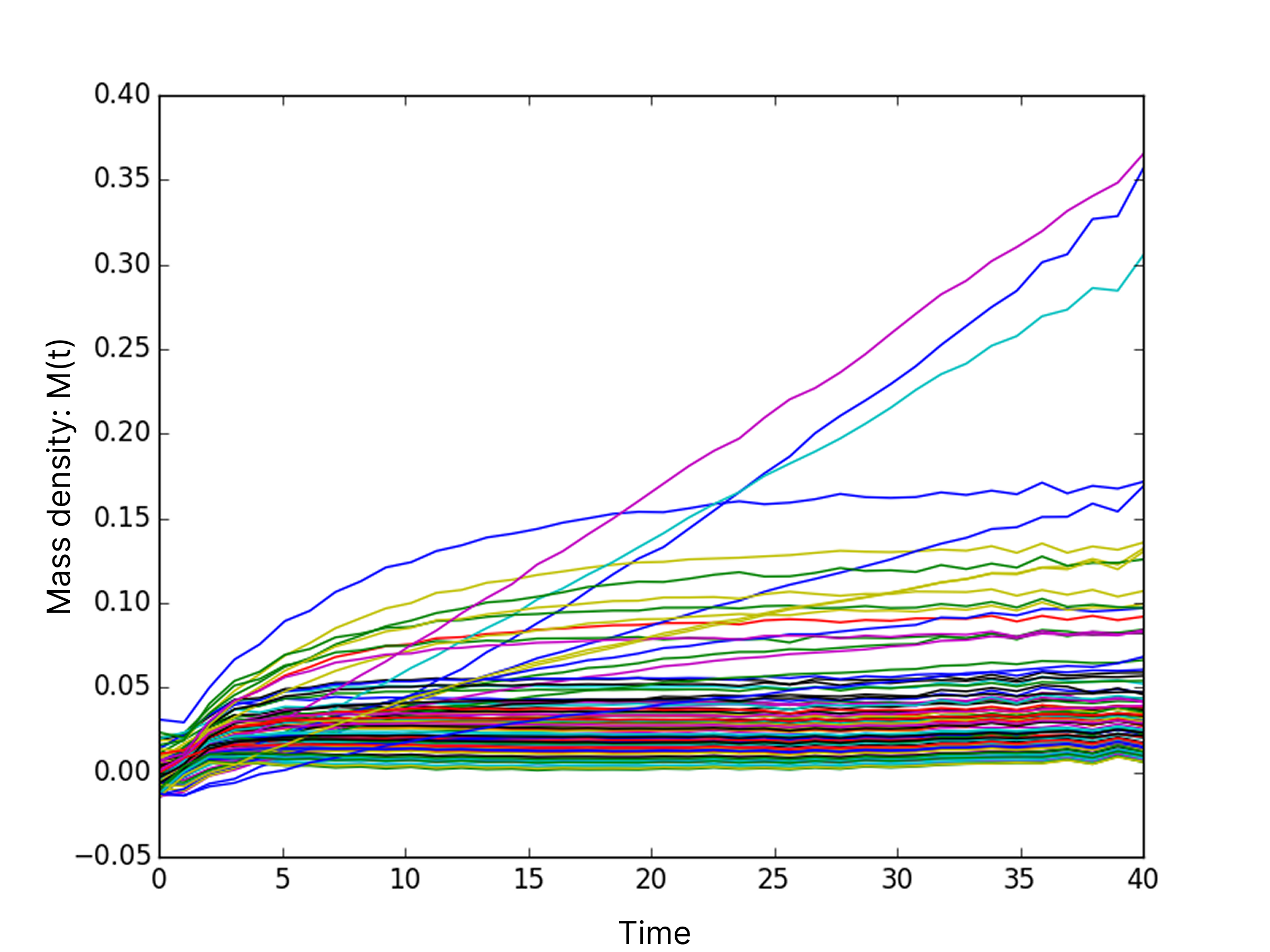}
        \caption{Reconstructed time series with convolutional output layer implemented}
        \label{fig:reconstruct_conv}
    \end{minipage}
\end{figure}

In order to test the above two methods, we here compared some preliminary results. Fig.~\ref{fig:orignal} is the original time series. Fig.~\ref{fig:reconstruct} is the reconstructed time series and due to lack of explicit modeling time dependency, although the magnitude of the reconstructed is correct, the reconstructed time series exhibit artificial non-smoothness. Fig.~\ref{fig:reconstruct_moving_average} shows the reconstructed time series with moving average method implemented. The reconstructed time series looks smoother and closer to the original time series. Fig.~\ref{fig:reconstruct_conv} is the results with one additional convolutional layer is added to the output layer. The reconstructed time series do look smoother but the trend and magnitude of the prediction is not as good as the results without convolutional layer. It is also easy to see that the prediction is not good on the boundary, which is due to padding for convolutional layer.

%\begin{figure}[!htb]
%    \centering
%    \begin{minipage}{.5\textwidth}
%        \centering
%        \includegraphics[width=1.0\textwidth]{M_moving_average_new}
%        \caption{Reconstructed time series with moving average implemented}
%        \label{fig:reconstruct_moving_average}
%    \end{minipage}%
%    \begin{minipage}{0.5\textwidth}
%        \centering
%        \includegraphics[width=1.0\textwidth]{M_reconstruct_conv_new}
%        \caption{Reconstructed time series with convolutional output layer implemented}
%        \label{fig:reconstruct_conv}
%    \end{minipage}
%\end{figure}

\section{Results}
The complete learning process to predict the time series $X$ based on patient-specific parameters $\theta$ has two stages. First an autoencoder is trained through all data points to obtain a compact (latent) representation $Z$ for each individual patient. Then a five-layer neural network is trained to obtain the mapping from patient-specific parameters $\theta$ to the latent representations $Z$. Additionally, physical model information can be imposed on the learning optimization problem for autoencoder as constraints.

\subsection{Reconstructed time series}
\begin{figure}
    \centering
    \begin{subfigure}{.6\textwidth}
        \centering
        \includegraphics[width=\textwidth]{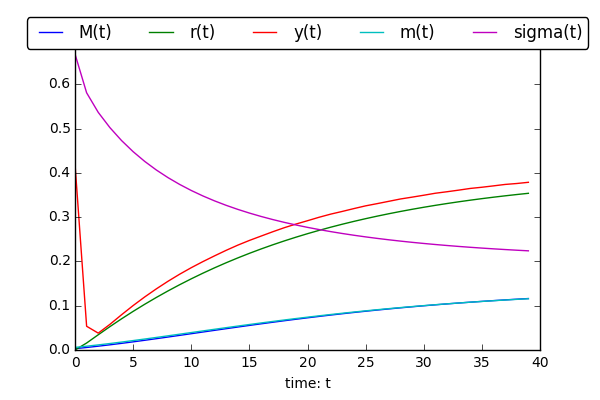}
        %\caption{Original time series for Patient 238 }
        %\label{fig:reconstruct_moving_average}
    \end{subfigure}%
\vfill
    \begin{subfigure}{0.6\textwidth}
        \centering
        \includegraphics[width=\textwidth]{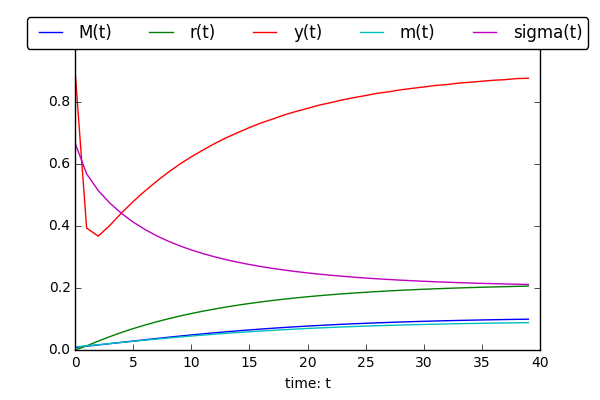}
        %\caption{Patient-238 reconstructed time series}
        %\label{fig:reconstruct_conv}
    \end{subfigure}
\vfill
    \begin{subfigure}{0.6\textwidth}
        \centering
        \includegraphics[width=\textwidth]{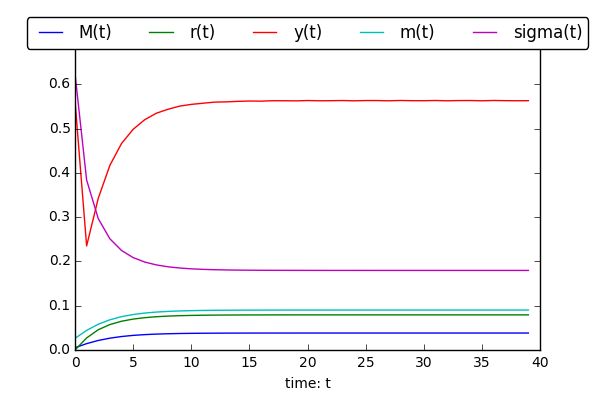}
        %\caption{Patient-238 reconstructed time series}
        %\label{fig:reconstruct_conv}
    \end{subfigure}
    \caption{The original time series of the five states $X(t) =[M(t), r(t), y(t), m(t), \sigma(t)]$ for Patient-238, Patient-315, Patient-2801}
    \label{fig:TS_original}
\end{figure}

\begin{figure}
    \centering
    \begin{subfigure}{.6\textwidth}
        \centering
        \includegraphics[width=\textwidth]{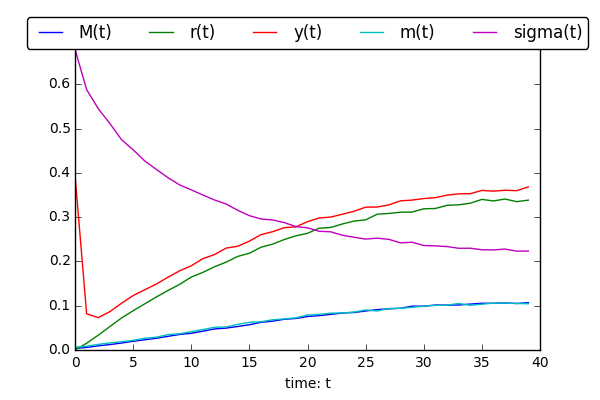}
        %\caption{Original time series for Patient 238 }
        %\label{fig:reconstruct_moving_average}
    \end{subfigure}%
\vfill
    \begin{subfigure}{0.6\textwidth}
        \centering
        \includegraphics[width=\textwidth]{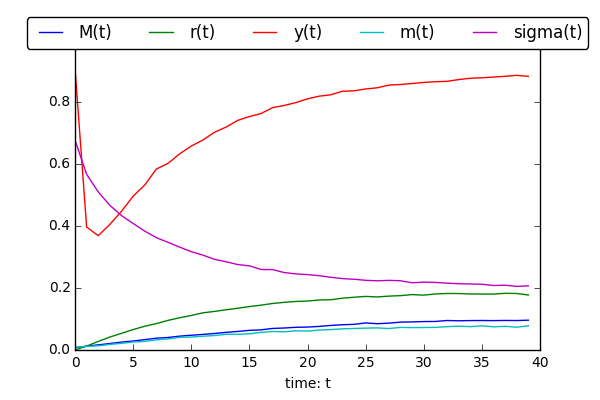}
        %\caption{Patient-238 reconstructed time series}
        %\label{fig:reconstruct_conv}
    \end{subfigure}
\vfill
    \begin{subfigure}{0.6\textwidth}
        \centering
        \includegraphics[width=\textwidth]{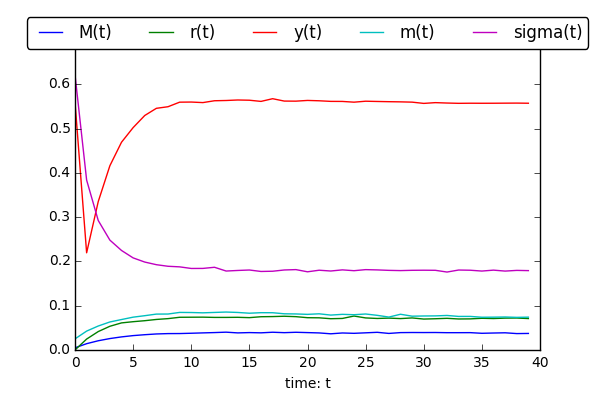}
        %\caption{Patient-238 reconstructed time series}
        %\label{fig:reconstruct_conv}
    \end{subfigure}
    \caption{The reconstructed time series of the five states $X(t) =[M(t), r(t), y(t), m(t), \sigma(t)]$ for Patient-238, Patient-315, Patient-2801}
    \label{fig:TS_reconstructed}
\end{figure}

\begin{figure}[!htb]
    \centering
    \begin{subfigure}{.8\textwidth}
        \centering
        \includegraphics[width=1.1\textwidth]{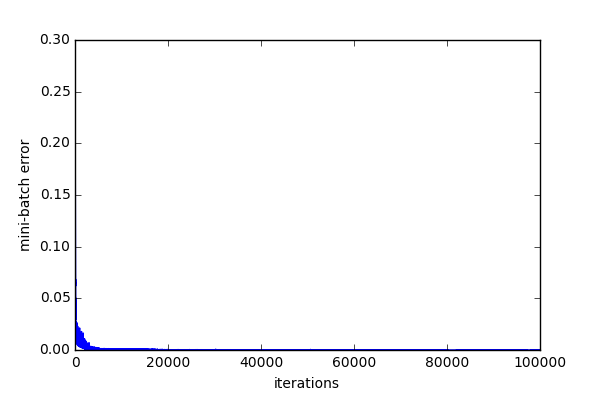}
        %\caption{Reconstructed time series with moving average implemented}
        %\label{fig:reconstruct_moving_average}
    \end{subfigure}%
\vfill
    \begin{subfigure}{0.8\textwidth}
        \centering
        \includegraphics[width=1.1\textwidth]{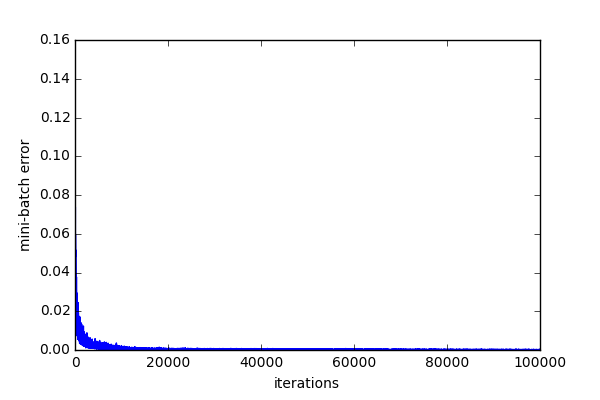}
        %\caption{Reconstructed time series with convolutional output layer implemented}
        %\label{fig:reconstruct_conv}
    \end{subfigure}
    \caption{Error convergence plots for the autoencoder (upper) and the neural network mapping from $\theta$ to $Z$ (lower)}
    \label{fig:error_convergence}
\end{figure}

The optimization problems in the above two stages are solved using stochastic gradient descent method~\cite{spall2005introduction}, with 2896 patient data sets for training and 100 data sets for testing. The convergence plots for the autoencoder and the mapping neural network are shown in Fig.~\ref{fig:error_convergence}. After training, given an new set of patient-specific parameters $\theta$, the latent representation $Z$ is first generated and then the time series $X$ are reconstructed using the decoder part of the autoencoder. Fig.~\ref{fig:TS_original} shows the true time series for three different patients (Patient-238, Patient-315, Patient-2801) with all the five states plotted in each  individual figure. Their time series have different dynamic behaviors due to different patient parameters. Fig.~\ref{fig:TS_reconstructed} shows the reconstructed time series for the three patients using our proposed prediction framework. It is easy to see that the reconstructed (predicted) time series match the true time series very well both in the trends and magnitudes of the time series. Note that these three patients are randomly selected from a test set that is different from the training set. Therefore, our model shows good performance in generalizing outside of the training set.  
%%%% original time series

%%%% reconstructed time series

%%%%%%% convergence plots

\subsection{Comparison between the results with and without constraints imposed}
\begin{figure}
    \centering
    \begin{minipage}{0.8\textwidth}
        \centering
        \includegraphics[width=1.0\textwidth]{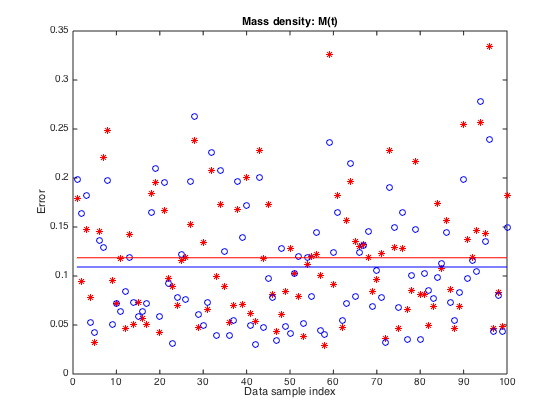}
        %\caption{Reconstructed time series with moving average implemented}
        %\label{fig:reconstruct_moving_average}
    \end{minipage}%
%\vfill
%    \begin{minipage}{0.33\textwidth}
%        \centering
%        \includegraphics[width=1.1\textwidth]{r}
%        %\caption{Reconstructed time series with convolutional output layer implemented}
%        %\label{fig:reconstruct_conv}
%    \end{minipage}
%    \begin{minipage}{0.33\textwidth}
%        \centering
%        \includegraphics[width=1.1\textwidth]{y}
%        %\caption{Reconstructed time series with convolutional output layer implemented}
%        %\label{fig:reconstruct_conv}
%    \end{minipage}
\vfill
    \begin{minipage}{0.8\textwidth}
        \centering
        \includegraphics[width=1.0\textwidth]{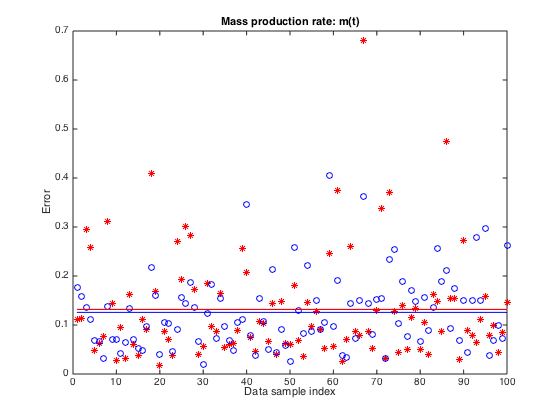}
        %\caption{Reconstructed time series with convolutional output layer implemented}
        %\label{fig:reconstruct_conv}
    \end{minipage}
%    \begin{minipage}{0.33\textwidth}
%        \centering
%        \includegraphics[width=1.1\textwidth]{sigma}
%        %\caption{Reconstructed time series with convolutional output layer implemented}
%        %\label{fig:reconstruct_conv}
%    \end{minipage}
    \caption{Reconstruction error plots for two of the states $[M(t), m(t)]$. (Red line) mean errors of all test patients without constraints. (Blue line) mean errors with constraints imposed.}
    \label{fig:error_no_noise}
\end{figure}

We first tested the reconstruction performance when there is no noise and bias in the input time series. The results are shown in Fig.~\ref{fig:error_no_noise}. It can be seen that imposing physical constraints does not necessarily help improve the prediction performance too much when the input data is without noise and bias. This is seems to be counterintuitive but actually reasonable. When the input data sets do not have noise and bias, the constraints have nothing to correct. The complexity of the neural networks is enough to capture the system dynamics of aneurysm growth. 

%Also adding constraints does not necessarily help to accelerate the convergence. This can be seen in the two special cases in Fig.~\ref{fig:two_special_cases}. The left figure in Fig.~\ref{fig:two_special_cases} shows that more intuitive case when the constraints help to reduce the searching space by projecting the solution trajectory onto the constrained subset. However, in other cases, usually when the constrained subset is non-convex, it may take more iteration steps by imposing the constraints (see Fig.~\ref{fig:two_special_cases}(right)).

\begin{figure}
    \centering
    \begin{minipage}{0.8\textwidth}
        \centering
        \includegraphics[width=1.0\textwidth]{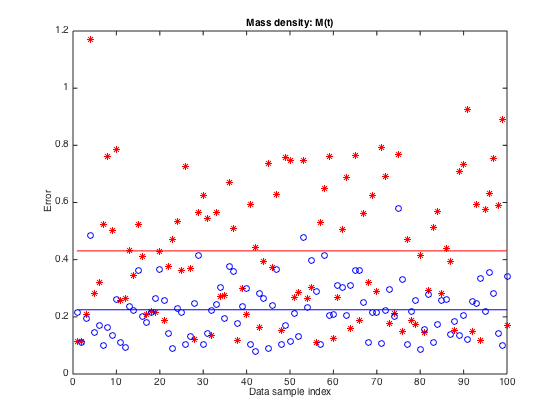}
        %\caption{Reconstructed time series with moving average implemented}
        %\label{fig:reconstruct_moving_average}
    \end{minipage}%
%    \begin{minipage}{0.19\textwidth}
%        \centering
%        \includegraphics[width=1.1\textwidth]{r_with_one_constraint}
%        %\caption{Reconstructed time series with convolutional output layer implemented}
%        %\label{fig:reconstruct_conv}
%    \end{minipage}
%    \begin{minipage}{0.19\textwidth}
%        \centering
%        \includegraphics[width=1.1\textwidth]{y_with_one_constraint}
%        %\caption{Reconstructed time series with convolutional output layer implemented}
%        %\label{fig:reconstruct_conv}
%    \end{minipage}
\vfill
    \begin{minipage}{0.8\textwidth}
        \centering
        \includegraphics[width=0.9\textwidth]{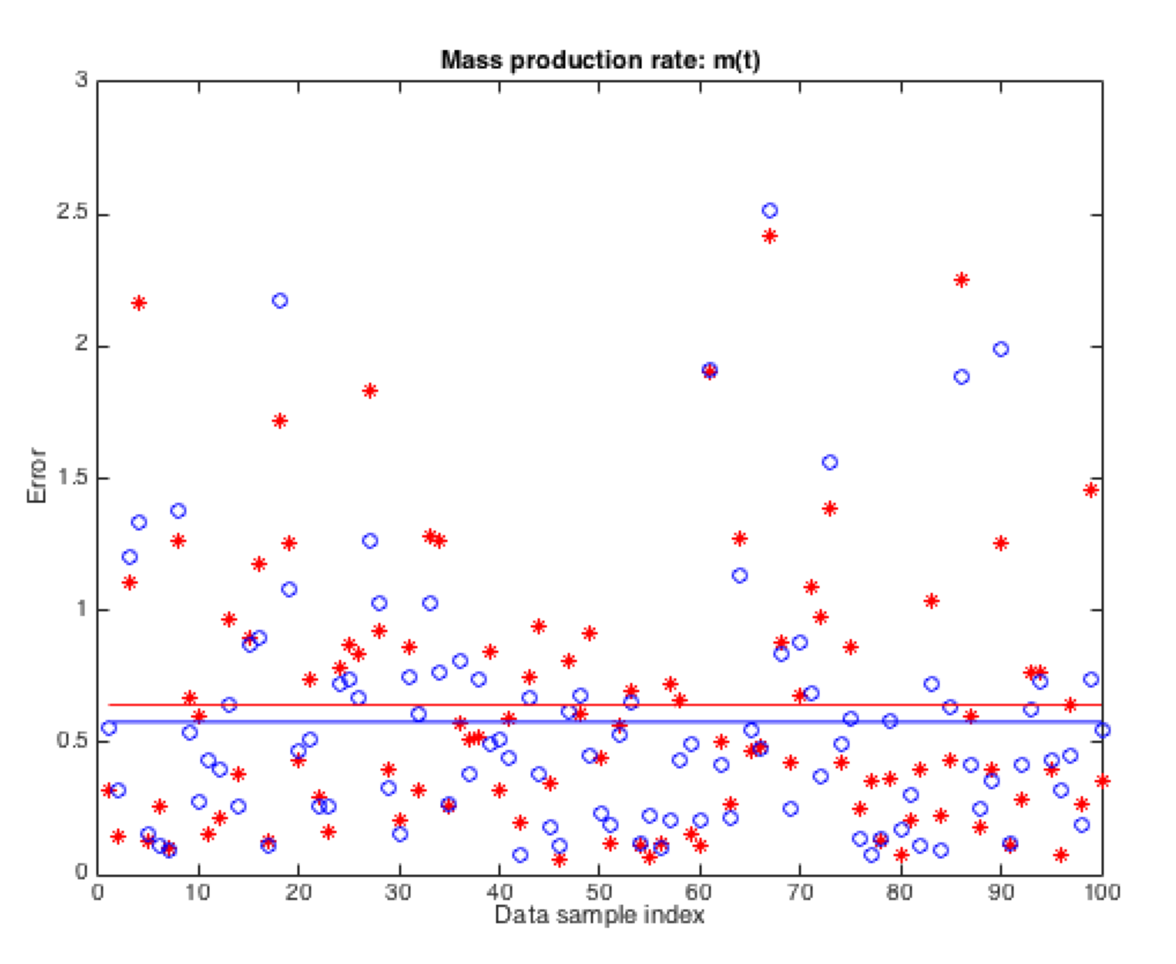}
        %\caption{Reconstructed time series with convolutional output layer implemented}
        %\label{fig:reconstruct_conv}
    \end{minipage}
%    \begin{minipage}{0.19\textwidth}
%        \centering
%        \includegraphics[width=1.1\textwidth]{sigma_with_one_constraint}
%        %\caption{Reconstructed time series with convolutional output layer implemented}
%        %\label{fig:reconstruct_conv}
%    \end{minipage}
    \caption{Reconstruction error plots for two of the states $[M(t), m(t)]$ with 10\% of random noise and 30\% of bias error.  (Red line) mean errors of all test patients without constraints. (Blue line) mean errors with one constraint $\sigma(t)=\frac{\rho P r(t)^2}{M(t) R}$ imposed.}
    \label{fig:one_constraint}
\end{figure}

\begin{figure}
    \centering
    \begin{minipage}{0.8\textwidth}
        \centering
        \includegraphics[width=1.0\textwidth]{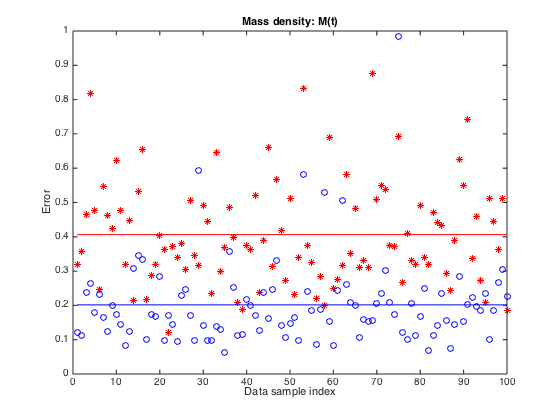}
        %\caption{Reconstructed time series with moving average implemented}
        %\label{fig:reconstruct_moving_average}
    \end{minipage}%
%    \begin{minipage}{0.19\textwidth}
%        \centering
%        \includegraphics[width=1.1\textwidth]{2r_with_two_constraints}
%        %\caption{Reconstructed time series with convolutional output layer implemented}
%        %\label{fig:reconstruct_conv}
%    \end{minipage}
%    \begin{minipage}{0.19\textwidth}
%        \centering
%        \includegraphics[width=1.1\textwidth]{2y_with_two_constraints}
%        %\caption{Reconstructed time series with convolutional output layer implemented}
%        %\label{fig:reconstruct_conv}
%    \end{minipage}
\vfill
    \begin{minipage}{0.8\textwidth}
        \centering
        \includegraphics[width=1.0\textwidth]{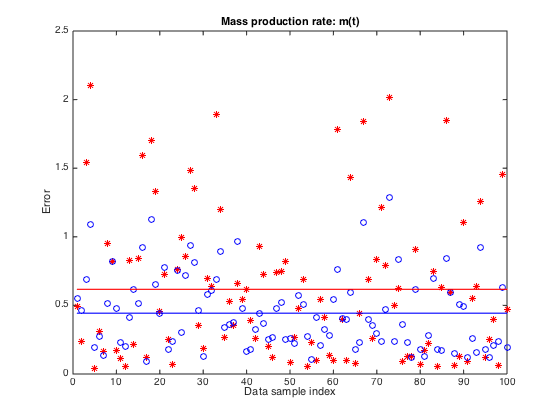}
        %\caption{Reconstructed time series with convolutional output layer implemented}
        %\label{fig:reconstruct_conv}
    \end{minipage}
%    \begin{minipage}{0.19\textwidth}
%        \centering
%        \includegraphics[width=1.1\textwidth]{2sigma_with_two_constraints}
%        %\caption{Reconstructed time series with convolutional output layer implemented}
%        %\label{fig:reconstruct_conv}
%    \end{minipage}
    \caption{Reconstruction error plots for two of the states $[M(t), m(t)]$ with 10\% of random noise and 30\% of bias error.  (Red line) mean errors of all test patients without constraints. (Blue line) mean errors with two constraints $\sigma(t)=\frac{\rho P r(t)^2}{M(t) R}$ and $m(t)=M(t)\left[k_g\left[\sigma(t)-\sigma_h\right] +f_h \right] $ imposed.}
    \label{fig:two_constraints}
\end{figure}

In order to prove that adding constraints can help to correct the prediction, we add 10\% of random noise and 30\% of bias error into the training data sets. We here compare the model prediction errors in two different cases: (a) with a constraint $\sigma(t)=\frac{\rho P r(t)^2}{M(t) R}$ imposed, (b) with two constraints $\sigma(t)=\frac{\rho P r(t)^2}{M(t) R}$ and $m(t)=M(t)\left[k_g\left[\sigma(t)-\sigma_h\right] +f_h \right] $ imposed. As we can see, in case (a) Fig.~\ref{fig:one_constraint}, the prediction error for $M(t)$ is significantly reduced  while errors for the other four variables do not change too much. This is because the given constraint $\sigma(t)=\frac{\rho P r(t)^2}{M(t) R}$ enforces a relation between $M(t)$ and $\sigma(t)$ which is proved to be the variable that is most accurately predicted. Therefore the constraint corrects the noise and bias in $M(t)$ and reduced the prediction error, as any solution that does not satisfy the constraint will be penalized.  In case (b) Fig.~\ref{fig:two_constraints}, both the errors for $M(t)$ and $m(t)$ are significantly reduced as apart from the reason mentioned before on why $M(t)$ becomes more accurate, the second constraint enforced a relation between $M(t)$ and $m(t)$, which helps to reduce the prediction error in $m(t)$.

%%%%%%% error without noise and bias

%%%%%% explain why adding constraints does not necessarily help faster convergence
%\begin{figure}
%    \centering
%        \begin{minipage}{0.49\textwidth}
%        \centering
%        \includegraphics[width=0.8\textwidth]{Constrained_opt_good}
%        %\caption{Reconstructed time series with convolutional output layer implemented}
%        %\label{fig:reconstruct_conv}
%    \end{minipage}
%    \begin{minipage}{.49\textwidth}
%        \centering
%        \includegraphics[width=0.8\textwidth]{Constrained_opt}
%        %\caption{Reconstructed time series with moving average implemented}
%        %\label{fig:reconstruct_moving_average}
%    \end{minipage}%
%    \caption{Left: the case when adding constraints help to reduce convergence steps. Right: the case when adding constraints does not help to reduce convergence steps. Note that blue lines are the convergence trajectory with constraints imposed and red lines are the convergence trajectory without constraints.}
%    \label{fig:two_special_cases}
%\end{figure}

%%%%%%% error with one constraint

%%%%%%% error with two constraints

\subsection{Vanishing gradient problem and normalization}
Vanishing gradient problem~\cite{hochreiter2001gradient} is an issue arisen when using back propagation method to train a neural network. This is due to the squashing property of the activation functions like sigmoid and $\tanh$ functions. This is in general unlikely to happen for a shallow five-layer neural network like in this paper. However it did show up when the input patient-specific parameters are not normalized. This is because the patient parameters $\theta$ has four different components with totally different order of magnitudes. This causes some of the input layer neurons directly enter into the saturation region and is never able to ecape from the dead zone during the whole training process. To address this, the patient-specific variables are normalized with respect to the nominal values to make the parameters unitless. Training with the unitless patient-specific parameters resolves the vanishing gradient problem.

\section{Conclusions}
In this paper, a learning framework is developed to reconstruct the arterial aneurysm growth time series from patient-specific parameters. The learning process consists of two stages: use autoencoder to obtain compact representations for the time series and then a five-layer neural network is trained to obtain the mapping from patient-specific parameters to the compact representations. Moving average method and convolutional output layer are implemented to account for the time dependency in the time series.

Physical model information is incorporated into the learning optimization problem as additional constraints for autoencoder training stage. The simulation results show that incorporating constraints does not help to improve performance or accelerate convergence if the date is noise and bias free. However, in the case of data with noise and bias, physics-informed learning (i.e. with constraints) can significantly reduce the prediction error as the physical constraints helped to correct the prediction of state variables against noise and bias. Considering measurement bias and noise universally exist in medical signals, adding constraints that represent physical models to the learning process will have a meaningful impact in various real-world applications.

%\section{others:}
%\begin{itemize}
%\item Maybe add things about stability analysis on the linear stability and how we choose the parameters
%\end{itemize}

\bibliographystyle{unsrt}
\bibliography{physical_learning}

\begin{thebibliography}{10}

\bibitem{cronenwett1985actuarial}
JL~Cronenwett, TF~Murphy, GB~Zelenock, Jr~WM Whitehouse, SM~Lindenauer,
  LM~Graham, LE~Quint, TM~Silver, and JC~Stanley.
\newblock Actuarial analysis of variables associated with rupture of small
  abdominal aortic aneurysms.
\newblock {\em Surgery}, 98(3):472--483, 1985.

\bibitem{humphrey2002constrained}
JD~Humphrey and KR~Rajagopal.
\newblock A constrained mixture model for growth and remodeling of soft
  tissues.
\newblock {\em Mathematical Models and Methods in Applied Sciences},
  12(03):407--430, 2002.

\bibitem{wu2015coupled}
J~Wu and SC~Shadden.
\newblock Coupled simulation of hemodynamics and vascular growth and remodeling
  in a subject-specific geometry.
\newblock {\em Annals of Biomedical Engineering}, 43(7):1543--1554, 2015.

\bibitem{willard2020integrating}
Jared Willard, Xiaowei Jia, Shaoming Xu, Michael Steinbach, and Vipin Kumar.
\newblock Integrating physics-based modeling with machine learning: A survey.
\newblock {\em arXiv preprint arXiv:2003.04919}, 1(1):1--34, 2020.

\bibitem{wu2019adding}
Jiacheng Wu, Jian-Xun Wang, and Shawn~C Shadden.
\newblock Adding constraints to bayesian inverse problems.
\newblock In {\em Proceedings of the AAAI Conference on Artificial
  Intelligence}, volume~33, pages 1666--1673, 2019.

\bibitem{raissi2019physics}
Maziar Raissi, Paris Perdikaris, and George~E Karniadakis.
\newblock Physics-informed neural networks: A deep learning framework for
  solving forward and inverse problems involving nonlinear partial differential
  equations.
\newblock {\em Journal of Computational physics}, 378:686--707, 2019.

\bibitem{boussel2008aneurysm}
Loic Boussel, Vitaliy Rayz, Charles McCulloch, Alastair Martin, Gabriel
  Acevedo-Bolton, Michael Lawton, Randall Higashida, Wade~S Smith, William~L
  Young, and David Saloner.
\newblock Aneurysm growth occurs at region of low wall shear stress:
  patient-specific correlation of hemodynamics and growth in a longitudinal
  study.
\newblock {\em Stroke}, 39(11):2997--3002, 2008.

\bibitem{wu2016stability}
Jiacheng Wu and Shawn~C Shadden.
\newblock Stability analysis of a continuum-based constrained mixture model for
  vascular growth and remodeling.
\newblock {\em Biomechanics and modeling in mechanobiology}, 15(6):1669--1684,
  2016.

\bibitem{ng2011sparse}
Andrew Ng.
\newblock Sparse autoencoder.
\newblock {\em CS294A Lecture notes}, 72(2011):1--19, 2011.

\bibitem{crank1947practical}
John Crank and Phyllis Nicolson.
\newblock A practical method for numerical evaluation of solutions of partial
  differential equations of the heat-conduction type.
\newblock In {\em Mathematical Proceedings of the Cambridge Philosophical
  Society}, volume~43, pages 50--67. Cambridge University Press, 1947.

\bibitem{galar2013aggregation}
Mikel Galar, Aranzazu Jurio, Carlos Lopez-Molina, Daniel Paternain, Jose Sanz,
  and Humberto Bustince.
\newblock Aggregation functions to combine rgb color channels in stereo
  matching.
\newblock {\em Optics express}, 21(1):1247--1257, 2013.

\bibitem{devcic2006weighted}
John Devcic.
\newblock Weighted moving averages: The basics, 2006.

\bibitem{lecun2015lenet}
Yann LeCun et~al.
\newblock Lenet-5, convolutional neural networks.
\newblock {\em URL: http://yann. lecun. com/exdb/lenet}, 2015.

\bibitem{spall2005introduction}
James~C Spall.
\newblock {\em Introduction to stochastic search and optimization: estimation,
  simulation, and control}, volume~65.
\newblock John Wiley \&amp; Sons, 2005.

\bibitem{hochreiter2001gradient}
Sepp Hochreiter, Yoshua Bengio, Paolo Frasconi, J{\"u}rgen Schmidhuber, et~al.
\newblock Gradient flow in recurrent nets: the difficulty of learning long-term
  dependencies, 2001.

\end{thebibliography}

\end{document}